\documentclass{optica-article}

\journal{opticajournal} 

\articletype{Research Article}

\usepackage{graphicx}
\usepackage{dcolumn}
\usepackage{bm}
\usepackage{braket}
\usepackage{mathcomp}

\usepackage{comment}
\usepackage{siunitx}
\usepackage{color}
 
\usepackage{here}

\begin{document}

\title{Quantum-enhanced optical phase-insensitive heterodyne detection beyond 3-dB noise penalty of image band}

\author{Keitaro Anai,\authormark{1} Yutaro Enomoto,\authormark{1,3} Hiroto Omura,\authormark{1} Koji Nagano,\authormark{2} Kiwamu Izumi,\authormark{2} Mamoru Endo,\authormark{1} and Shuntaro Takeda\authormark{1,4}}

\address{\authormark{1}Department of Applied Physics, School of Engineering, The University of Tokyo, 7-3-1 Hongo, Bunkyo-ku, Tokyo 113-8656, Japan\\
\authormark{2}Institute of Space and Astronautical Science (JAXA), Chuo-ku, Sagamihara City, Kanagawa 252-5210, Japan}

\email{\authormark{3}email: yenomoto.ap.t@gmail.com}
\email{\authormark{4}email: takeda@ap.t.u-tokyo.ac.jp}


\begin{abstract*} 

Optical phase-insensitive heterodyne (beat-note) detection, which measures the relative phase of two beams at different frequencies through their interference, is a key sensing technology for various spatial/temporal measurements, such as frequency measurements in optical frequency combs. However, its sensitivity is limited not only by shot noise from the signal frequency band but also by the extra shot noise from an image band, known as the 3-dB noise penalty. Here, we propose a method to remove shot noise from all these bands using squeezed light. We also demonstrate beyond-3-dB noise reduction experimentally, confirming that our method actually reduces shot noise from both the signal and extra bands simultaneously. Our work should boost the sensitivity of various spatial/temporal measurements beyond the current limitations.

\end{abstract*}

\section{Introduction}
{
Optical phase-insensitive heterodyne detection (also known as beat-note detection), which measures the relative phase of two monochromatic beams at different frequencies through their interference, is a key sensing technology for various spatial/temporal ultra-precise measurements (this is in contrast to the phase-sensitive heterodyne detection to measure a single quadrature through the interference of monochromatic and bichromatic beams). For example, such detection is essential for high-precision clock comparisons \cite{ths:clock}, frequency measurements in optical frequency combs \cite{ths:comb, ths:comb_make}, and space gravitational wave telescopes \cite{ths:graviton}. Furthermore, heterodyne detection is generally advantageous when low-frequency optical signals contain important information \cite{ths:graviton, ths:decigo, ths:gravity_sq, ths:beat_thermal}. This is because it can detect low-frequency optical signals as high-frequency electronic signals and thus has immunity to common low-frequency electronic noise~\cite{ths:hetero}.
}

\begin{figure}[htbp]
    \centering
    \includegraphics[width=7cm]{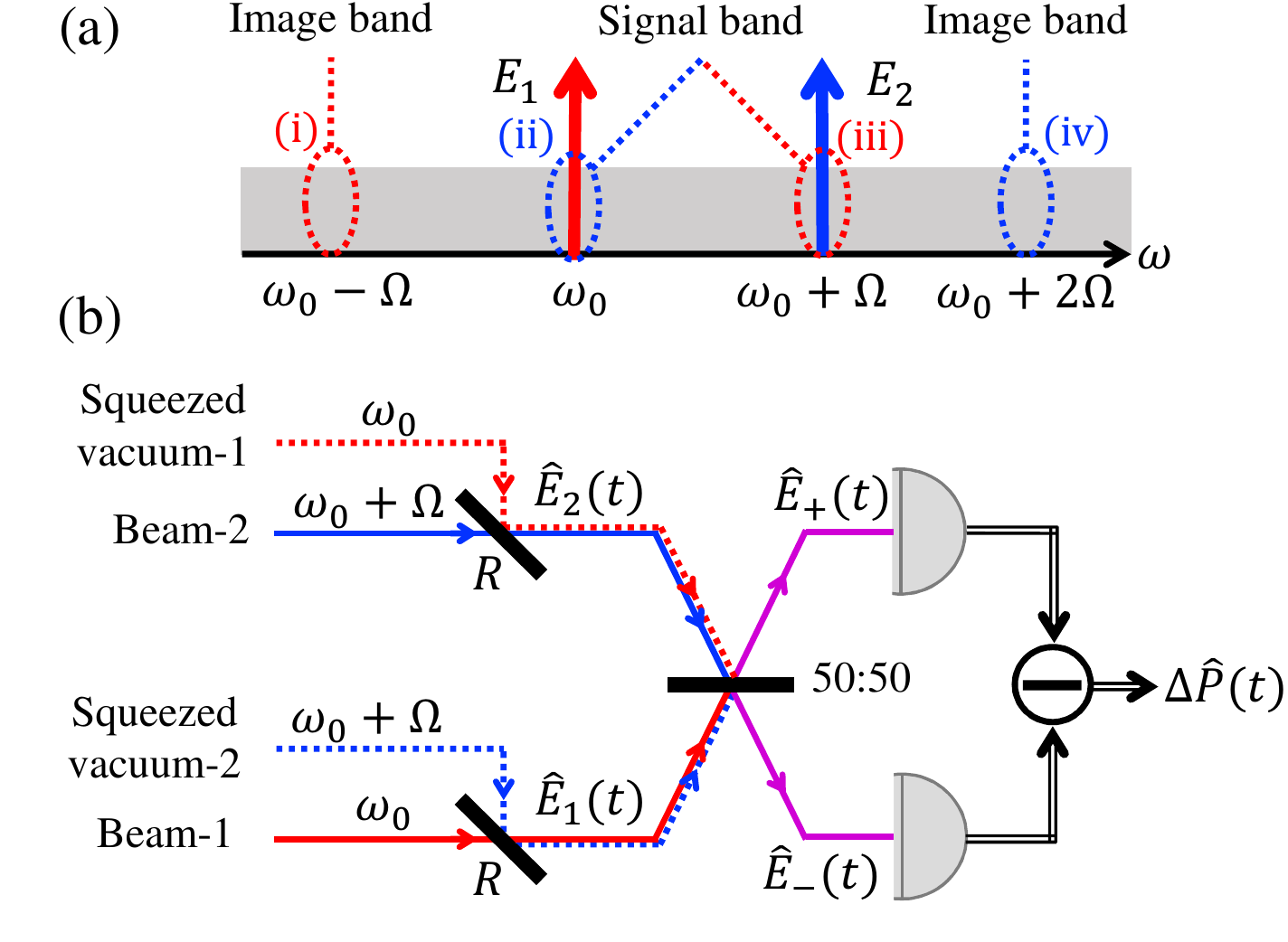}
    \caption{
Optical phase-insensitive heterodyne detection. (a) Signals and quantum fluctuation in the frequency domain. Solid arrows represent classical amplitudes, while the grey-shaded area represents quantum fluctuation. Dotted circles represent the quantum fluctuation affecting phase-insensitive heterodyne detection. (b) Conceptual schematic. Squeezed vacuum-1 (2) makes a quantum correlation between (i) and (iii) ((ii) and (iv)) in (a).
       }
    \label{fig:theory}
    \end{figure}

{
The sensitivity of phase-insensitive heterodyne detection, like other optical sensing methods, is fundamentally limited by noise due to the quantization of light, commonly known as shot noise~\cite{ths:shot_theory, ths:b}. In fact, this detection is inherently more susceptible to shot-noise limitations than other methods due to the notorious ``extra noise penalty," which doubles the contribution of shot noise~\cite{ths:shot_theory, ths:b}. This penalty comes from the contamination by the image-band shot noise when measuring signals in the signal band (Fig.\ref{fig:theory}(a)). Thus, the sum of the contributions from the signal and image bands determines the classical sensitivity limit of phase-insensitive heterodyne detection. In general, the effects of shot noise in optical sensing can be reduced by increasing the optical power. However, this strategy is also disadvantageous for phase-insensitive heterodyne detection because it unavoidably uses large classical beat signals, which make optical detectors saturated or even broken. This can be compared to, for example, homodyne detection, which can extract phase information without large signals by using dark fringe conditions \cite{ths:enomoto_suggestion}.
}

It has long been known that shot noise can be reduced by using squeezed light even under optical power constraints~\cite{ths:sq_review, ths:sq_bio, ths:magnet_sq}. Such an approach has been proposed and demonstrated in single-quadrature measurement by homodyne detection~\cite{ths:sq_old_2} with a monochromatic local oscillator (LO) or phase-sensitive heterodyne detection with a bichromatic LO~\cite{ths:hetero,ths:hetero_2,ths:a,ths:d,ths:e,ths:g, ths:Rev_0_1, ths:EPR_3}. However, its application to phase-insensitive heterodyne detection, which uses only a single sideband instead of symmetric sidebands as in the phase-sensitive one and measures both phase and amplitude quadratures at the same time, has been limited because the simultaneous reduction of all of the signal- and image-band noise seems nontrivial. The use of the squeezed light for the signal band~\cite{ths:sq_old_2, ths:hetero, ths:hetero_2} can only reduce half shot noise and leave image-band noise intact, limiting the noise reduction below 3 dB. Ref.\cite{ths:c} has reported squeezing-based enhancement of phase-insensitive heterodyne detection in a limited situation when one of the two beams (signal) is much weaker than the other (LO). However, it adopts a non-universal detection system combining squeezed LO generation, signal-beam damping, and signal-LO mixing, and beyond-3-dB noise reduction has yet to be achieved even in this limited situation.

{
Here, we propose and demonstrate a general method to remove shot noise from both the signal and image bands in phase-insensitive heterodyne detection (Fig.\ref{fig:theory}(b)). The key idea is to use squeezed light of two different frequencies, correlate quantum fluctuation in the signal and image bands, and thereby cancel out their effect at the same time. In principle, our method can completely remove shot noise in the limit of infinite squeezing. We experimentally implement this method and reduce shot noise by more than 3 dB at $\sim$3 MHz sideband frequency around the 10 MHz beat signal. This beyond-3-dB noise reduction is unachievable only with the signal-band squeezing, and thus clearly demonstrates the shot-noise reduction from both the signal and image bands. Our method can be applied to various applications of phase-insensitive heterodyne detection, and boost the sensitivity beyond the current limitations.
}

\section{Working principle}

{
In general, optical phase-insensitive heterodyne detection is used to extract the information encoded in the relative phase of two beams at different frequencies. This detection can be described as follows 
(Fig.\ref{fig:theory}(a)). First, two incoming beams at angular frequencies of $\omega_0$ (beam-1) and $\omega_0+\Omega$ (beam-2) interfere with each other. The optical power after the interference beats at the angular frequency $\Omega$. By tracking the phase of this beat note, we can continuously measure the relative phase of the initial two beams. 
}

{
The sensitivity of the phase-insensitive heterodyne detection depends on the signal-to-noise ratio of the phase component of the beat signal at $\Omega$. This signal is contaminated by quantum fluctuation at optical angular frequencies $\omega_0-\Omega$, $\omega_0$, $\omega_0+\Omega$, and $\omega_0+2\Omega$ because these angular frequencies are $\Omega$ away from the angular frequency of beam-1 or beam-2. The angular frequencies around $\omega_0$ and $\omega_0+\Omega$ are called ``signal bands," and their quantum fluctuation is inseparable from the phase signal of interest. On the other hand, the angular frequencies around $\omega_0-\Omega$ and $\omega_0+2\Omega$ are called ``image bands," and their quantum fluctuation also contributes to noise even though these bands have no signal of interest. Therefore, the classical sensitivity limit of phase-insensitive heterodyne detection is determined by the collective contributions from both the signal and image bands, and the contribution of the image-band vacuum in this detection is 
often called the extra noise penalty of 3 dB.
}

{
A typical way to reduce shot noise is to use squeezed lights. The straightforward idea is to inject phase-squeezed light to beam-1 (2) so that these two have the same center angular frequency $\omega_0$ ($\omega_0+\Omega$). As a result, we can suppress the quantum fluctuation of the phase component at the corresponding signal band $\omega_0$ ($\omega_0+\Omega$) while not affecting other bands as long as the bandwidth of the squeezed light is appropriate. In this case, however, quantum fluctuation at the image bands remains (See Appendix G for details).
}

{
In contrast, our method can simultaneously reduce shot noise in both the signal and image bands. The key idea is to inject squeezed light to beam-1 and 2 in such a way that the frequencies are inversely related; that is, we inject the squeezed light at $\omega_0+\Omega$ ($\omega_0$) to beam-1 (2) with an appropriate phase. The squeezed light makes a quantum correlation between quantum fluctuation at $\omega_0$ and $\omega_0+2\Omega$ ($\omega_0-\Omega$ and $\omega_0+\Omega$) \cite{ths:two_photon}. When this light interferes with beam-2 (1) and is detected, the quantum fluctuation is canceled and reduced due to the quantum correlation. As a result, we can simultaneously suppress the shot noise of both the signal and image bands, and reach beyond-3-dB shot noise reduction. For this purpose, the squeezed light can be injected in a way shown in Fig.\ref{fig:theory}(b). This phase-insensitive heterodyne detection assumes a situation with optical power constraints, where the beams are picked off by beam splitters with high reflectivity $R$ to set their power below the upper limit. For example, such a situation includes the case when the beam with a signal needs to be attenuated before the detector to avoid its saturation or breakdown. In this case, we can inject squeezed lights through the other ports of the beam splitters. The squeezed light suffers a $(1-R)$ optical loss, which can be made arbitrarily small if sufficiently large optical power is initially available for beam-1 and beam-2. Alternatively, the squeezed light can also be injected without strongly attenuating beam-1 and beam-2 by using optical cavities as dichroic filters~\cite{ths:FilterCavity}. This is possible because the squeezed light and the beam to be combined have different frequencies in our scheme. This strategy is useful to enhance the sensitivity even further when the available beam power is limited.
}

{
Next, we introduce a mathematical formulation of the principle. Here we adopt the two-photon formalism~\cite{ths:two_photon} because it can conveniently describe squeezed light's entanglement, which exists between two sidebands equally separated from the center frequency. According to the two-photon formalism, we can define sideband operators with center angular frequency $\omega$ and sideband angular frequency $\epsilon$ in the frequency domain as $\hat{a}_1^\omega(\epsilon)=(\hat{a}_{\omega+\epsilon}+\hat{a}^\dagger_{\omega-\epsilon})/\sqrt{2}$, $\hat{a}_2^\omega(\epsilon)=(\hat{a}_{\omega+\epsilon}-\hat{a}^\dagger_{\omega-\epsilon})/i\sqrt{2}$. Here, $\hat{a}_\omega$ is a photon annihilation operator of angular frequency $\omega$. We can then define the operators for quadrature phase amplitudes in the time domain as $\hat{a}_j^\omega(t)=\int_0^\infty\mathrm{d}\epsilon[\hat{a}_j^\omega(\epsilon)e^{-i\epsilon t}+\{\hat{a}_j^{\omega}(\epsilon)\}^\dagger e^{i\epsilon t}]/2\pi$, where $t$ denotes time and $j=1,2$.
In Fig.\ref{fig:theory}(b), beam-1 is first attenuated at the beam splitter with high reflectivity $R$. The amplitude of beam-1 before the beam splitter is sufficiently large and chosen as $E_1e^{i\omega_0t}/\sqrt{1-R}$, in which case beam-1 can be treated as classical and its quantum fluctuation can be ignored. At this attenuation, quantum fluctuation comes in from the other port of the beam splitter. It can be expressed as $(\hat{a}_1^{\omega_0+\Omega}(t)-i\hat{a}_2^{\omega_0+\Omega}(t))e^{i(\omega_0+\Omega)t}$, where the center angular frequency ($\omega_0+\Omega$) is intentionally shifted from that of beam-1 ($\omega_0$) for later convenience. From these definitions, the beam after the high-reflectivity beam splitter can be described as
\begin{equation}
    \hat{E}_1(t)=\frac{E_1e^{i\omega_0t}}{\sqrt{1-R}}\cdot\sqrt{1-R}+(\hat{a}_1^{\omega_0+\Omega}(t)-i\hat{a}_2^{\omega_0+\Omega}(t))e^{i(\omega_0+\Omega)t}\cdot\sqrt{R},\label{eq:BeforePickoff}
\end{equation}
and in the limit of $R\rightarrow1$, we obtain
\begin{equation}
    \hat{E}_1(t)=E_1e^{i\omega_0t}
    +(\hat{a}_1^{\omega_0+\Omega}(t)-i\hat{a}_2^{\omega_0+\Omega}(t))e^{i(\omega_0+\Omega)t}.\label{eq:field1}
\end{equation}
In the same way, the beam after the attenuation of beam-2 can be described as
\begin{equation}
    \hat{E}_2(t)=E_2e^{i((\omega_0+\Omega)t+\theta(t))}+(\hat{b}_1^{\omega_0}(t)-i\hat{b}_2^{\omega_0}(t))e^{i\omega_0t},\label{eq:field2}
\end{equation}
\noindent where $(\hat{b}_1^{\omega_0}(t)-i\hat{b}_2^{\omega_0}(t))e^{i\omega_0t}$ represents quantum fluctuation from the other port than beam-2. The goal of the phase-insensitive heterodyne detection is to measure the relative phase $\theta(t)$ of these two beams. The 50:50 beam splitter transforms these beams into $\hat{E}_\pm(t)=(\hat{E}_1(t)\pm\hat{E}_2(t))/\sqrt{2}$. We then measure their beam-power difference. The measured signal $\Delta \hat{P}(t)=\hat{E}_+^\dagger(t) \hat{E}_+(t)-\hat{E}_-^\dagger(t) \hat{E}_-(t)$ can be described as
\begin{align}
\Delta \hat{P}(t)=&2[E_1E_2\cos(\Omega t +\theta(t)) \nonumber\\
&+E_2\hat{a}_1^{\omega_0+\Omega}(t)+E_1\hat{b}_1^{\omega_0}(t)],
\label{eq:output}
\end{align}
\noindent by using Eqs.~(\ref{eq:field1}) and (\ref{eq:field2}) while neglecting the second-order terms for the quantum fluctuation and $\theta(t)$. Here, we assume that $\theta(t)$ is sufficiently smaller than 1. The explicit derivation of Eq.~\eqref{eq:output} under the more general case where $\theta(t)$ is not small is described in Appendix F. {In Eq.~\eqref{eq:output}, $E_1E_2\cos(\Omega t +\theta(t))$ represents the classical beat signal, and $E_2\hat{a}_1^{\omega_0+\Omega}(t)+E_1\hat{b}_1^{\omega_0}(t)$ represents the shot noise which limits the sensitivity of the phase-insensitive heterodyne detection. In our method, both of the shot-noise terms $\hat{a}_1^{\omega_0+\Omega}(t)$ and $\hat{b}_1^{\omega_0}(t)$ can be arbitrarily reduced by appropriately adjusting the relative phase between squeezed vacuum-1 and beam-1
(squeezed vacuum-2 and beam-2) so that the noise component $\hat{a}_1^{\omega_0+\Omega}(t)$ ($\hat{b}_1^{\omega_0}(t)$) is squeezed as long as the half-width of half maximum of the squeezing is larger than $\Omega$, respectively. Thus, Eq.(\ref{eq:output}) shows that we can reduce the shot-noise contributions to zero in the limit of infinite squeezing. Note that the degree of the shot-noise suppression is independent of $E_1$ and $E_2$ as long as the squeezing levels of the term $\hat{a}_1^{\omega_0+\Omega}(t)$ and $\hat{b}_1^{\omega_0}(t)$ are the same.
 }

 {
The squeezing of $\hat{a}_1^{\omega_0+\Omega}(t)$ ($\hat{b}_1^{\omega_0}(t)$) is also equivalent to canceling out shot noise at $\omega_0$ and $\omega_0+2\Omega$ ($\omega_0-\Omega$ and $\omega_0+\Omega$) by making an Einstein-Podolsky-Rosen (EPR) state~\cite{ths:EPR_1, ths:EPR_2, ths:EPR_3} in these angular frequencies. This can be understood from the identity (Appendix A)
\begin{align}\label{eq:EPR_relation}
2\hat{a}_1^{\omega_0+\Omega}(t)=&
(\hat{a}_2^{\omega_0+2\Omega}(t)-\hat{a}_2^{\omega_0}(t))\sin\Omega t \nonumber \\
&+(\hat{a}_1^{\omega_0+2\Omega}(t)+\hat{a}_1^{\omega_0}(t))\cos\Omega t,
\end{align}
\noindent
where both the first and second terms of the right-hand side are reduced for the EPR state (the same discussion holds for $\hat{b}_1^{\omega_0}(t)$).
For this reason, our method can be interpreted as the use of a quantum correlation between signal and image bands to reduce noise.
 }

{
 Equation~(\ref{eq:output}) shows that two conjugate observables (amplitude and phase) of the beat signal at $\Omega$ can be simultaneously measured below the vacuum-noise level, seemingly violating the uncertainty principle. However, this is not the case since our situation can be interpreted as one of the realizations of the quantum dense metrology \cite{ths:metro} by using the EPR state in the frequency domain. The similar tricks can also be seen in Refs.~\cite{ths:uncertain_1,ths:uncertain_2,ths:uncertain_3,ths:uncertain_4}.
}


\section{Experimental setup}
\begin{figure}[htbp]
    \centering
    \includegraphics[width=13cm]{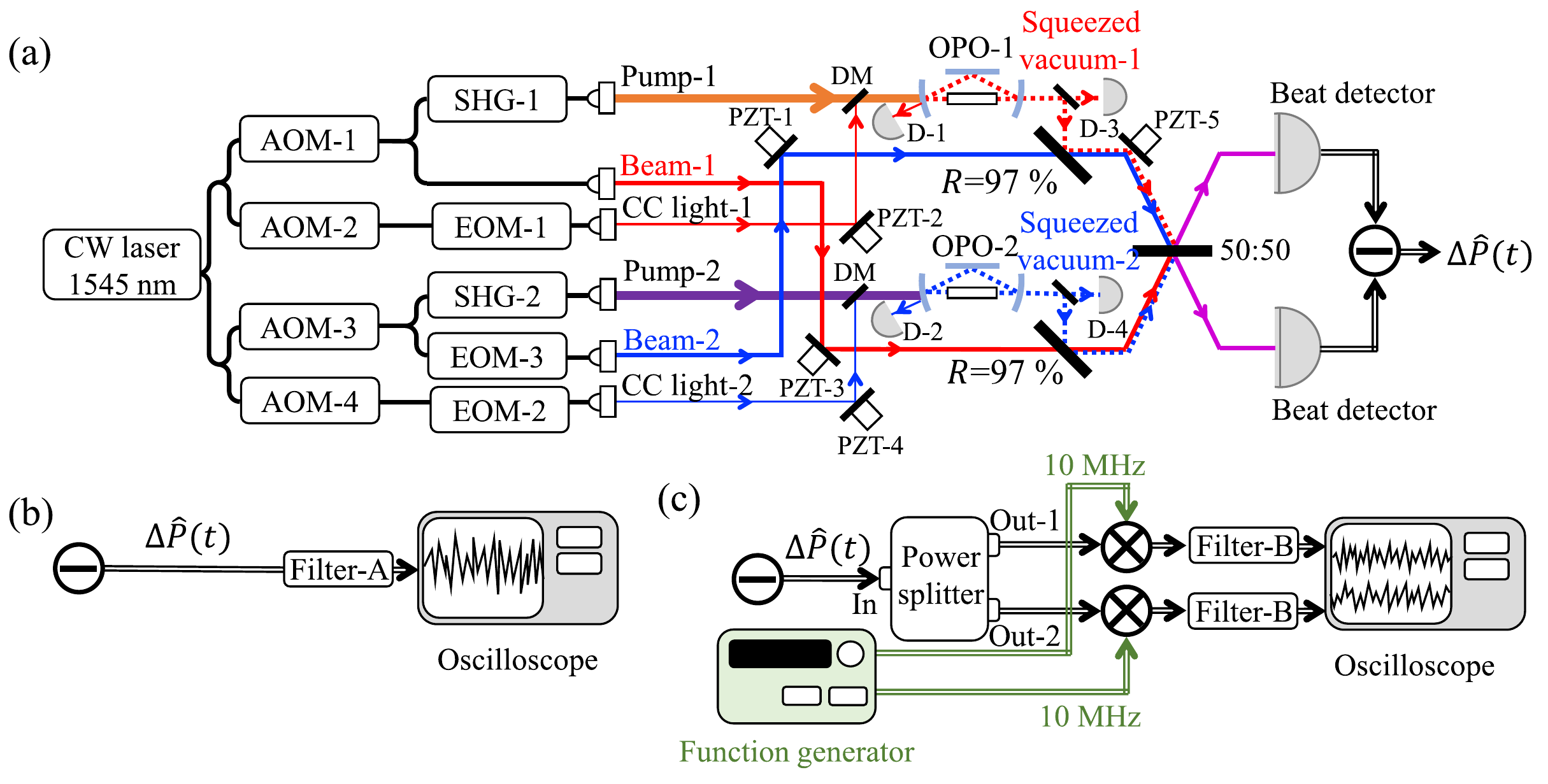}
    \caption{{{Experimental setup for phase-insensitive heterodyne detection}}
            {(a)} {Optical setup. CW, continuous-wave; AOM, acousto-optic modulator; SHG, second harmonic generator; OPO, optical parametric oscillator; EOM, electro-optic modulator; DM, dichroic mirror; CC light, classical coherent light; PZT, piezo actuator. D-$i$ $(i=1,\cdots,4)$ are photo detectors.}
            {(b)} {Electronic setup for raw-signal measurement.}
            {(c)} {Electronic setup for demodulated-signal measurement.}
        }
    \label{fig:experiment}
\end{figure}

{
Figure~\ref{fig:experiment}(a) shows our experimental setup. We use a continuous-wave (CW) laser at 1545 nm (NKT Photonics, Koheras ADJUSTIK \& BOOSTIK). The beat frequency is set to 10 MHz and created by the interference of beam-1 and 2, whose frequencies are downshifted by 85 MHz and 75 MHz from that of the source laser with acousto-optic modulators-1 and -3 (AOM, Gooch \& Housego, Fiber-Q), respectively. The power of both beams is set to 0.6 mW before the pickoff beam splitter in order to suppress the unwanted effect of laser intrinsic phase noise derived from laser line width (line width < 15 kHz; laser intrinsic phase-noise at 1-m optical path difference < 36 $\mathrm{\mu}$rad/$\sqrt{\mathrm{Hz}}$ at 100 Hz and 3.5 $\mathrm{\mu}$rad/$\sqrt{\mathrm{Hz}}$ at 20 kHz, according to the specification sheet), etc. while obtaining the measurable shot-noise power (See Appendix C for more details about the choice of the beam power). Beam-2 is phase-modulated at 3.11 MHz by an electro-optic modulator-3 (EOM, iXblue, MPZ-LN-01). This modulation mimics the relative-phase signal in the phase-insensitive heterodyne detection and corresponds to $\theta(t)$ in Eq.(\ref{eq:field1}). In our experiment, we choose the modulation frequency as 3.11 MHz because this frequency domain does not contain excess experimental noise and is also within the bandwidth of the OPOs.
}

{
Two squeezed lights are injected at the beam splitters with $97\%$ reflectivity to reduce shot noise in this detection. These lights are produced by two triangle-shaped optical parametric oscillators (OPO) with a PPKTP crystal inside. These OPOs have a half width at the half maximum of 30 MHz, which sufficiently covers the beat frequency of 10 MHz (similar design as in Ref.~\cite{ths:OPO}). The oscillation thresholds of these OPOs are estimated to be $\sim$600 mW. The cavity lengths of OPO-1 and OPO-2 are locked by modulating classical coherent (CC) light-1 and -2 with EOM-1 and -2 and detecting them with detectors D-1 and -2, respectively. The pump beams for the OPOs, pump-1 and 2, are downshifted by $85\times2$ MHz and $75\times2$ MHz from the second harmonic beam of the source laser and set at $\sim$90 mW and $\sim$80 mW, which is far below the oscillation threshold. Pump-1 and pump-2 are generated by doubling the frequency with a second harmonic generator (SHG, NTT Electronics, WH-0773-000-F-B-C)-1 and -2 after downshifting the source laser by $85$ MHz and $75$ MHz with an AOM (Gooch \& Housego, Fiber-Q)-1 and -3, respectively.
}

{
To lock the phase relationship, the beat lock (coherent control) technique~\cite{ths:CoherentControl} is used by additionally injecting CC lights into the OPOs, as shown in Fig.~\ref{fig:experiment}(a). CC light-1 (CC light-2) is downshifted by 84.65 MHz (74.74 MHz) from the source laser with AOM (Gooch \& Housego, Fiber-Q)-2 and -4, and set to $\sim0.2\ \mathrm{\mu W}$ before the ``beat detector'' shown in Fig.~\ref{fig:experiment}(a). First, we lock the relative phase between squeezed vacuum-1 and CC light-1 (squeezed vacuum-2 and CC light-2) by partially detecting the signal of parametrically amplified CC light-1 (CC light-2) with detector D-3 (D-4) and demodulating it at 700 kHz (520 kHz). The demodulated signal is fed back to PZT-2 (PZT-4). Next, we lock the relative phase between CC light-1 and beam-1 (CC light-2 and beam-2) by obtaining their interference signal with the beat detector and demodulating it at 350 kHz (260 kHz). The signal is fed back to PZT-5 (PZT-1). As a result of these locking, the relative phase between squeezed vacuum-1 and beam-1 (squeezed vacuum-2 and beam-2) are locked. Our phase-locking system is sufficiently stable to always keep the phases locked during the measurement time. These CC lights are turned off during data acquisition to avoid additional noise by employing the sample-and-hold method \cite{ths:sample}. In a single cycle (500 $\mathrm{\mu s}$) of the sample-and-hold method, 400 $\mathrm{\mu s}$ is used to actively lock the phases with the CC lights (sample time), while 100 $\mathrm{\mu s}$ is used to hold the phases and acquire data without the CC lights (hold time). The sample time is taken sufficiently short for the phases to be stable for this measurement.
}

{
We detect the beat signal by a $\sim$ 90-MHz-bandwidth homemade beat detector with photodiodes of 99\% quantum efficiency (Laser Components). The gain flatness of the beat detector is below 1 dB among the frequencies used in this experiment (within $\pm$5 MHz from the 10-MHz beat signal). As already described above, beam-1 and beam-2 are set to 0.6 mW before the high-reflectivity beam splitter, which is far below the saturation power of the beat detector. Under this condition, the clearance of the vacuum shot noise power to the electronic noise power is about 2 dB.

With this beat detector, we acquire the signal in two ways as shown in Figs.~\ref{fig:experiment}(b) and (c). First, we directly acquire the raw beat signal (Fig.~\ref{fig:experiment}(b)) as the simplest setting. In this case, the sum of amplitude and phase noise can be analyzed. Second, we acquire the demodulated signal (Fig.~\ref{fig:experiment}(c)) to analyze only phase noise. At the demodulation, the relative phase of the 10-MHz beat note and the 10-MHz local oscillator signal is locked to $\pi/2$ by feeding the demodulated signal back to PZT-3 as an error signal. In this case, we can obtain only the phase noise because the amplitude component of the beat note becomes out-of-phase with the local oscillator and disappears after the demodulation, while the phase component becomes in-phase with the local oscillator and survives after the demodulation. In addition to this configuration, we use the cross-spectrum method to reduce electronic noise~\cite{ths:cross_spec}. In this case, a power splitter splits the raw signal into two, which are then separately demodulated by 10-MHz local oscillator signals generated from separate channels of a function generator. We finally take a cross-spectrum of these two demodulated signals to reduce electronic noise: we first acquire the temporal waveforms of both signals ($v_1(t)$, $v_2(t)$), perform the Fourier transform of them with a Hamming window ($V_1(\omega)$, $V_2(\omega)$), compute their product after taking the complex conjugate of one of them ($V_1(\omega)V_2^\ast(\omega)$), and take its real part. Repeating the above procedure and averaging the cross-spectrum over sufficient time suppress independent noise generated from electronic components after the power splitter. We use an oscilloscope (bandwidth: 1 GHz, sampling rate: 125 MHz) because it enables us to conveniently perform all the necessary data acquisition and post-processing for both the raw-signal and demodulated-signal measurements described above. Several electronic filters are used to exploit the full dynamic range of the oscilloscope (Appendix B).
}

\section{Experimental results}

{
As a preliminary measurement, we evaluate the squeezing level of squeezed vacuum-1 (2) using beam-1 (2) as a local oscillator for homodyne detection. During the measurement, beam-2 (1) and squeezed vacuum-2 (1) are blocked. The squeezing level of squeezed vacuum-1 (2) at lower and upper sidebands are 4.5(1) dB and 3.7(1) dB (4.16(8) dB and 4.0(1) dB), respectively (See Appendix E for more details about the preliminary evaluation of our OPOs). Here, the lower and upper sidebands correspond to the frequency range of $\pm0.5$ MHz around $10-3.11$ MHz and $10+3.11$ MHz, respectively. The squeezing levels of OPO-1 and -2 are almost the same because we use the symmetric setup as Fig.~\ref{fig:experiment}(a) shows. These squeezing levels are limited by the optical loss of $20\%$ (estimated from the asymmetry between the squeezing and anti-squeezing level) and the pump power limitation due to the tolerable optical power of acousto-optic modulators for the pump beams. This estimated loss is reasonably well explained by $\sim$4\% internal loss of the OPOs, $\sim$6\% optical propagation loss (including $3\%$ loss of the $97\%$-reflectivity beam splitter), $\sim$5\% spatial mode mismatch between the beams, $\sim$1\% beat detector inefficiency, and the other small unmeasurable experimental imperfections. The shot-noise reduction in the phase-insensitive heterodyne detection should reflect these initial squeezing levels.
}

{
Next, the shot-noise reduction in the phase-insensitive heterodyne detection is evaluated by taking three types of data for both the raw and demodulated measurements. The first is the data of the background electronic noise without any light (background). The second is the data of the unsqueezed shot noise with only beam-1 and beam-2 (reference). The third is the data of the squeezed shot noise with the squeezed lights in addition to beam-1 and beam-2 (target). For each data, we obtain the power spectrum by averaging over 12500 frames of a time-series waveform consisting of 5000 sampling points. In our setting, the power of shot noise to be measured is comparable to that of electronic noise. To analyze the pure noise reduction of shot noise itself, we subtract the power spectrum of electronic noise (background) from those of unsqueezed and squeezed shot noise (reference and target). Then we plot these power spectra after normalizing them by setting the average of the pure unsqueezed-shot-noise level to 0 dB. The details of such post-processing are described in Appendix C.
}

{
Figure \ref{fig:raw_result} shows the spectra of the raw measurement around the lower and upper sidebands of the 10-MHz beat signal after subtraction of the electric noise. The phase modulation of beam-2 appears as two sideband peaks at $10\pm 3.11$ MHz around the 10-MHz beat signal, and the noise floor around these peaks reflects the sum of amplitude and phase noise. In this case, the shot-noise level with squeezed light is reduced by 3.71(4) dB and 3.39(8) dB around the lower and upper sidebands, respectively. Here, the noise levels are calculated by averaging over the frequency range of $\pm0.5$ MHz around the peaks after excluding that of $\pm0.04$ MHz. The uncertainties are calculated from the standard error of data points in the frequency ranges described above. This result indicates that the reduction in the sum of amplitude and phase noise surpasses the 3-dB noise penalty.
}

\begin{figure}[H]
    \centering
    \includegraphics[width=12cm]{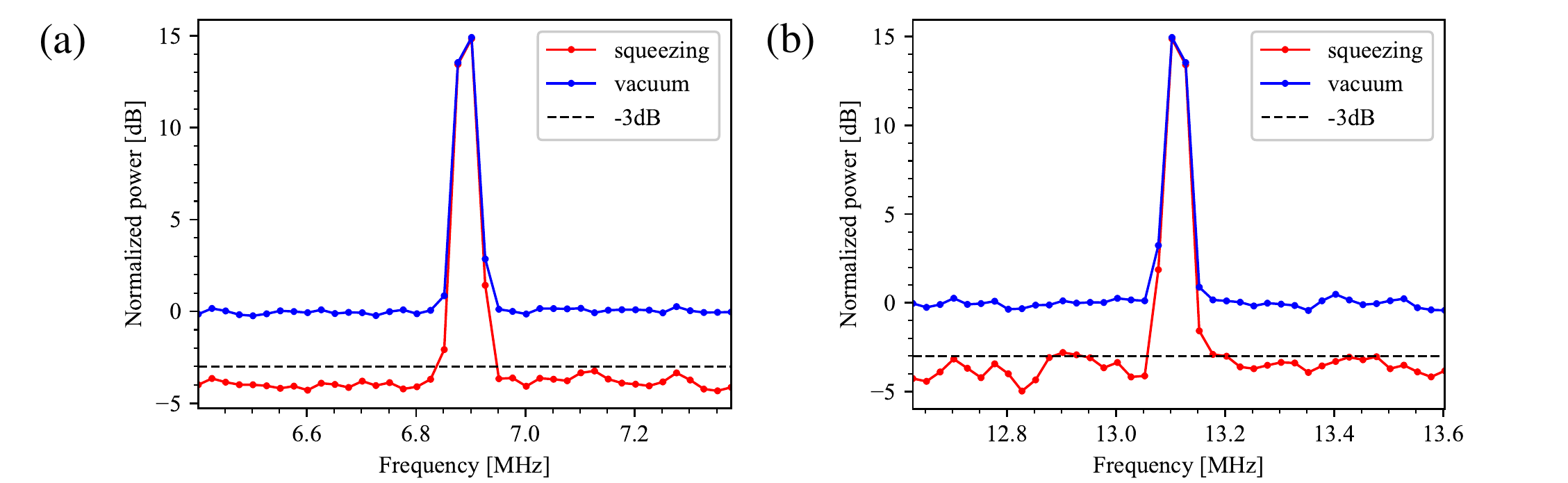}
    \caption{Experimental results of raw-signal measurement. The electronic noise was 2 dB below shot noise and subtracted from all traces. (a) Spectrum for the lower sideband at $10-3.11$ MHz. (b) Spectrum for the upper sideband at $10+3.11$ MHz. In both cases, the shot-noise level with squeezed light (red line) is lower than that with vacua (blue line) by more than the extra noise penalty of 3 dB (black dotted line). }
    \label{fig:raw_result}
\end{figure}

{
Figure \ref{fig:demod} shows the spectra of the demodulated measurement. After the demodulation, two sideband peaks at $10\pm 3.11$ MHz in Fig.~\ref{fig:raw_result} are converted in frequency and appear as a single peak at $3.11$ MHz, and the noise floor around it reflects only phase noise. In this case, the shot-noise level with squeezed light is reduced by 3.35(6) dB. Here, the noise levels are calculated by averaging over the frequency range of $\pm0.5$ MHz around the peak after excluding that of $\pm0.05$ MHz. The uncertainties are calculated from the standard error of data points in the frequency ranges described above. This result indicates that the reduction in phase noise alone surpasses the 3-dB noise penalty.
}

\begin{figure}[htbp]
    \centering
    \includegraphics[width=7cm]{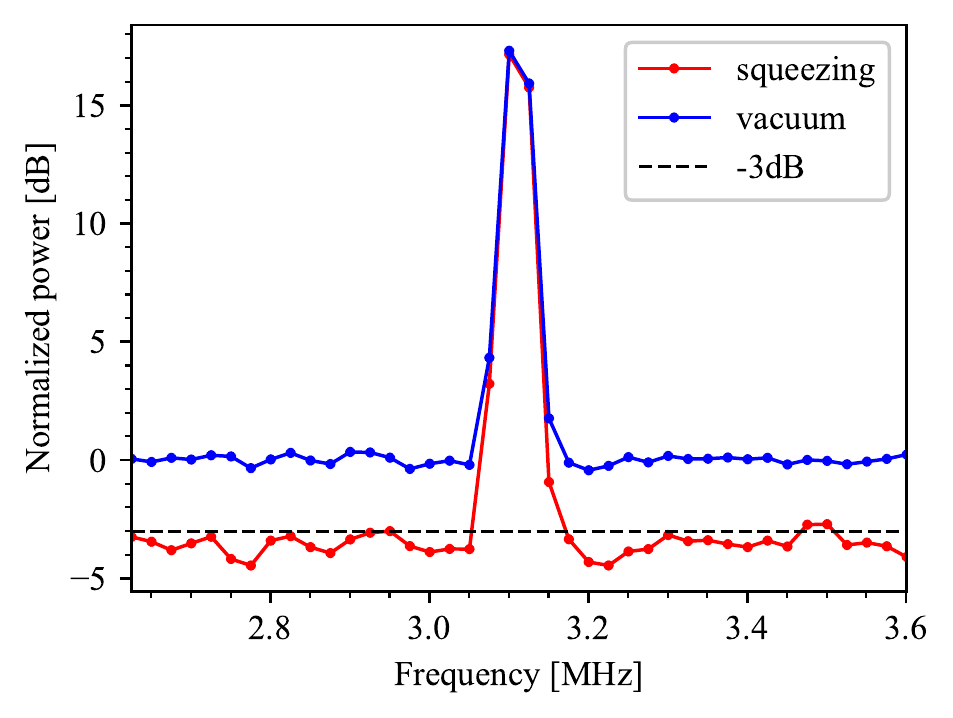}
    \caption{Experimental results of demodulated-signal measurement. The electronic noise was 2 dB below shot noise and subtracted from all traces. The shot-noise level with squeezed light (red line) is lower than that with vacua (blue line) by more than the extra noise penalty of 3 dB (black dotted line). 
        }
    \label{fig:demod}
\end{figure}

{
Note that the noise-reduction levels expected from the initial squeezing levels of squeezed vacuum-1 and 2 are 4.3 dB (3.8 dB) for the lower (upper) sideband in the raw measurement and 4.1 dB for the demodulated measurement. The actual noise-reduction levels in the raw and demodulated measurements are comparable but slightly smaller than these expected levels. This difference may be attributed to laser intrinsic phase noise of beam-1 and beam-2 that can be added to shot noise. We estimate that approximately $10 \%$ of the shot-noise power in Figs.~\ref{fig:raw_result} and \ref{fig:demod} can come from laser intrinsic phase noise. Under this condition, the noise reduction becomes 9.3 dB if the state-of-the-art 15 dB squeezing is available~\cite{ths:15dBsqueezer}, and is limited to 10 dB even with infinite squeezing. This means that our method is more effective for experimental systems and applications where classical noise is sufficiently small compared to shot noise. We also note that the impurity of the squeezed states only affects the final noise reduction in our phase-insensitive heterodyne detection as the decrease in the effective squeezing level, and the anti-squeezing level does not affect the detection in principle.
}

\section{Conclusion}

We proposed and theoretically proved a method to reduce shot noise in phase-insensitive heterodyne detection to zero using squeezed light with infinite squeezing level (Eq.(\ref{eq:output})). We also experimentally demonstrated that our method reduced the shot-noise level by more than the extra noise penalty of 3 dB in both raw and demodulated measurements of the beat signal. The noise-level reduction ($<$4 dB) and applicable frequency range (a few tens of MHz) are limited in our specific setup, but can be straightforwardly improved by replacing our squeezers with high-level (up to 15 dB~\cite{ths:15dBsqueezer}) or broader-bandwidth (over THz~\cite{ths:OPA}) ones. The phase-insensitive heterodyne detection in various applications is often limited by shot noise under some optical power constraints, and thus our method is effective for further reducing the noise level. It can be employed in various systems such as high-precision clock comparisons \cite{ths:clock}, frequency measurements in optical frequency combs \cite{ths:comb, ths:comb_make}, and space gravitational wave telescopes \cite{ths:graviton}.

During the revision process of this article, we have become aware of a recent work~\cite{ths:h2} demonstrating quantum-enhanced phase-insensitive heterodyne detection. While Ref.~\cite{ths:h2} assumed a limited situation where a weak signal beam is measured by a strong LO and reported beyond 3-dB noise reduction by a single squeezed light for such a highly asymmetric situation, our work covers a more general situation where two beams are strong by removing not only (i) and (iii) but also (ii) and (iv) in Fig.\ref{fig:theory}(a) with a pair of squeezed lights.

\begin{backmatter}

\bmsection{Appendix A. Relation between squeezing and quantum correlation}

Here we derive Eq.~\eqref{eq:EPR_relation} in the main text. As is described in the main text, we define sideband operators with center angular frequency $\omega$ and sideband angular frequency $\epsilon$ in the frequency domain as $\hat{a}_1^\omega(\epsilon)=(\hat{a}_{\omega+\epsilon}+\hat{a}^\dagger_{\omega-\epsilon})/\sqrt{2}$, $\hat{a}_2^\omega(\epsilon)=(\hat{a}_{\omega+\epsilon}-\hat{a}^\dagger_{\omega-\epsilon})/i\sqrt{2}$. Here, $\hat{a}_\omega$ is a photon annihilation operator of angular frequency $\omega$. We can then define the operators for quadrature phase amplitudes in the time domain as $\hat{a}_j^\omega(t)=\int_0^\infty\mathrm{d}\epsilon[\hat{a}_j^\omega(\epsilon)e^{-i\epsilon t}+\{\hat{a}_j^{\omega}(\epsilon)\}^\dagger e^{i\epsilon t}]/2\pi$, where $t$ denotes time and $j=1,2$.

First, the term $\hat{a}_1^{\omega_0+\Omega}(t)$ can be written as
\begin{equation*}
    \hat{a}_1^{\omega_0+\Omega}(t)=(\hat{a}_1^{\omega_0+\Omega}(t)\sin\Omega t)\sin\Omega t+(\hat{a}_1^{\omega_0+\Omega}(t)\cos\Omega t)\cos\Omega t.
\end{equation*}
With the definitions of $\hat{a}_j^{\omega}(t)$ and $\hat{a}_j^{\omega}(\epsilon)$ described in the main text and the equation $\{\hat{a}_j^\omega(-\epsilon)\}^\dagger=\hat{a}_j^\omega(\epsilon)\ (j=1,2)$,  we can transform the term $\hat{a}_1^{\omega_0+\Omega}(t)\sin\Omega t$ into
\begin{equation*}
    \begin{split}
        \hat{a}_1^{\omega_0+\Omega}(t)\sin\Omega t&=\int_0^\infty\frac{\mathrm{d}\epsilon}{2\pi}\left(\hat{a}_1^{\omega_0+\Omega}(\epsilon)e^{-i\epsilon t}+\{\hat{a}_1^{\omega_0+\Omega}(\epsilon)\}^\dagger e^{i\epsilon t}\right)\sin\Omega t\\
        &=\int_{-\infty}^\infty\frac{\mathrm{d}\epsilon}{2\pi}\hat{a}_1^{\omega_0+\Omega}(\epsilon)e^{-i\epsilon t}\sin\Omega t\\
        &=\int_{-\infty}^\infty\frac{\mathrm{d}\epsilon}{4\sqrt{2}\pi i}\left(\hat{a}_{\omega_0+\Omega+\epsilon}+\hat{a}_{\omega_0+\Omega-\epsilon}^\dagger\right)\left(e^{-i(\epsilon-\Omega)t}-e^{-i(\epsilon+\Omega)t}\right)\\
        &=\int_{-\infty}^\infty\frac{\mathrm{d}\epsilon}{4\sqrt{2}\pi i}\left[\left(\hat{a}_{\omega_0+2\Omega+\epsilon}+\hat{a}_{\omega_0-\epsilon}^\dagger\right)-\left(\hat{a}_{\omega_0+\epsilon}+\hat{a}_{\omega_0+2\Omega-\epsilon}^\dagger\right)\right]e^{-i\epsilon t}\\
        &=\int_{-\infty}^\infty\frac{\mathrm{d}\epsilon}{4\sqrt{2}\pi i}\left[\left(\hat{a}_{\omega_0+2\Omega+\epsilon}-\hat{a}_{\omega_0+2\Omega-\epsilon}^\dagger\right)-\left(\hat{a}_{\omega_0+\epsilon}-\hat{a}_{\omega_0-\epsilon}^\dagger\right)\right]e^{-i\epsilon t}\\
        &=\frac{1}{2}\left(\hat{a}_2^{\omega_0+2\Omega}(t)-\hat{a}_2^{\omega_0}(t)\right).
    \end{split}
\end{equation*}
In the same way, we can also derive
\begin{equation*}
    \hat{a}_1^{\omega_0+\Omega}(t)\cos\Omega t=\frac{1}{2}\left(\hat{a}_1^{\omega_0+2\Omega}(t)+\hat{a}_1^{\omega_0}(t)\right).
\end{equation*}
Hence, we derived Eq.~\eqref{eq:EPR_relation} in the main text.

\bmsection{Appendix B.Electrical filters in measurements}

\begin{figure}[htbp]
    \centering
    \includegraphics[width=12cm]{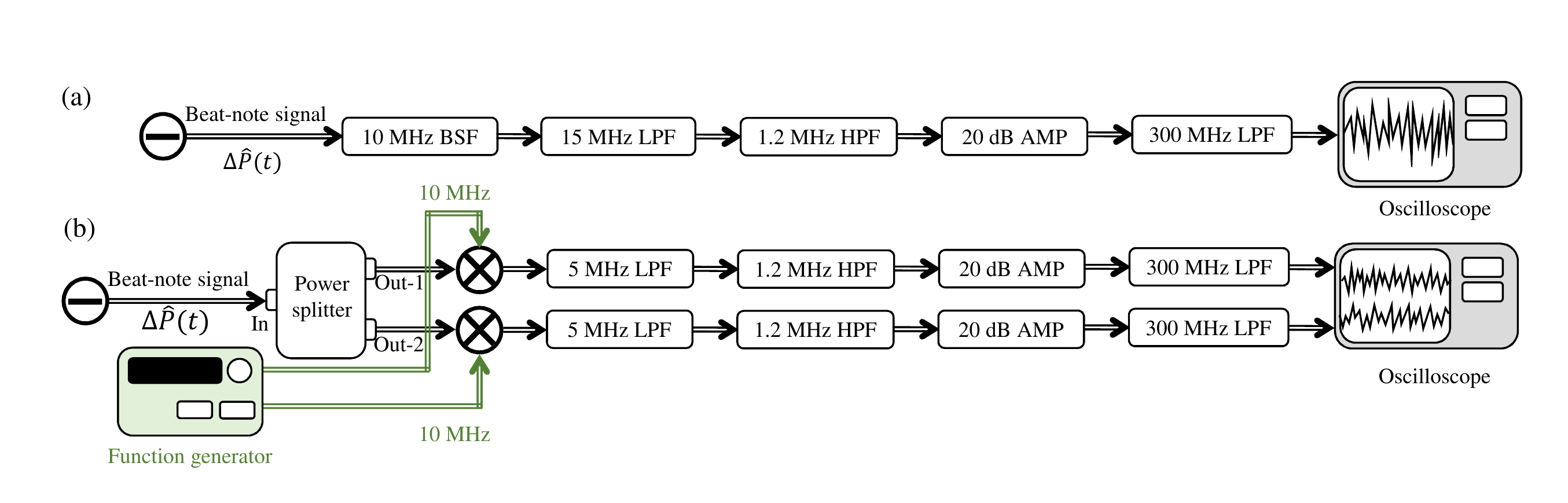}
    \caption{Filters used in our experiment. BSF, band stop filter; LPF, low pass filter; HPF, high pass filter; AMP, amplifier. See text for details. (a) A set of filters used in the raw-signal measurement. (b) A set of filters used in the demodulated-signal measurement.}
    \label{fig:filters}
\end{figure}

\noindent
As described in the main text, we implement both the raw- and demodulated-signal measurements for phase-insensitive heterodyne detection. Different filters are used for these two measurements to cut unnecessary peaks and amplify only the signal of interest while avoiding saturation of electrical circuits. Below, we describe the details of the filters.

The filters used in the raw-signal measurement are shown in {Fig.\ref{fig:filters}(a)}. The raw signal in phase-insensitive heterodyne detection first passes through a 10-MHz band-stop filter (homemade, 5th-order Chebyshev band-stop filter) to suppress the beat signal at 10 MHz and avoid saturation on the oscilloscope. The signal passes through a 15-MHz low-pass filter (Mini-circuits, BLP-15+) to suppress the unnecessary peaks at higher frequencies such as the second harmonic of the beat signal at 10 MHz, and it passes through a 1.2-MHz high-pass filter (Mini-circuits, ZFHP-1R2-S+) to suppress signals at lower frequencies used to lock the system, then being amplified by a 20-dB amplifier (homemade). Finally, the signal passes through a 300-MHz low-pass filter (Mini-circuits, BLP-300+) to further suppress higher frequency components.

The filters used in the demodulated-signal measurement are shown in {Fig.\ref{fig:filters}(b)}. In this case, the raw signal in phase-insensitive heterodyne detection is first divided into two by a power splitter (Mini-circuits, ZSC-2-1+) to take the cross-spectrum in the last step and thereby suppress the extra electric noise after the power splitter. These two signals are separately demodulated by mixers (Mini-circuits, ZAD-3+) with 10-MHz local oscillator signals. These 10-MHz signals are synchronized in phase but generated independently using the different channels of a function generator. Each of the demodulated signals passes through a 5-MHz low-pass filter (Mini-circuits, BLP-5+) to suppress the unnecessary signals at higher frequencies such as the 10-MHz peak, and it passes through a 1.2-MHz high-pass filter, a 20-dB amplifier, and a 300-MHz low-pass filter. The types and roles of these filters are the same as those used in the raw-signal measurement.

As a preliminary measurement, the frequency response of the entire set of filters is measured for each case of the raw- and demodulated-signal measurements. We use the measured response to compensate for the distortion of the spectra obtained in our phase-insensitive heterodyne detection. The power and cross-spectra in Figs.~3 and 4 in the main text are plotted after this compensation.

\bmsection{Appendix C. Wide-frequency-range spectra without post-processing}
\begin{figure}[htbp]
    \centering
    \includegraphics[width=12cm]{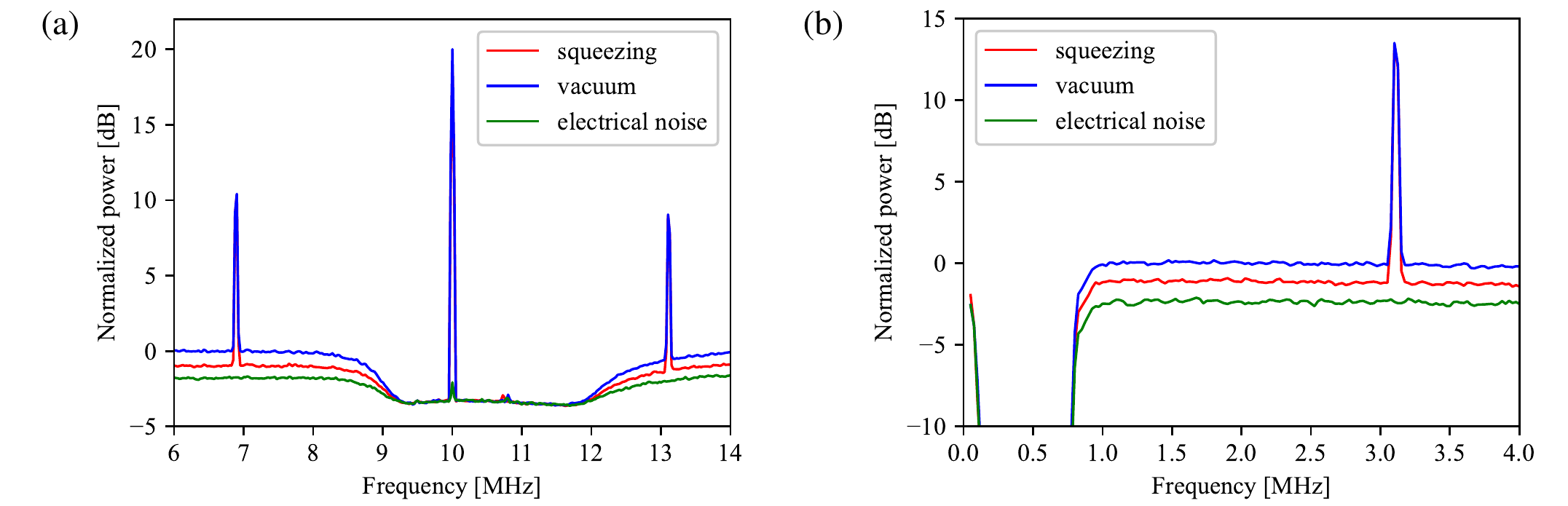}
    \caption{The broad spectra without post-processing. (a) A spectrum in the raw-signal measurement. The spectrum is normalized by the power of the vacuum averaging from 6 MHz to 6.5 MHz. (b) A 
    spectrum in the demodulated-signal measurement. The spectrum is normalized by the power of the vacuum averaging from 1 MHz to 3 MHz.}
    \label{fig:broad_spectra}
\end{figure}

\noindent
In the main text, we only showed the post-processed narrow-frequency-range spectra near the modulation peaks as the results of the phase-insensitive heterodyne detection. The post-processing includes the subtraction of the power of the background electrical noise as well as the compensation of the spectrum distortion by the electrical filters. For reference, here we plot the wide-frequency range spectrum without post-processing for both the raw- and demodulated-signal measurements in Fig.~\ref{fig:broad_spectra}. Each plot has three lines: the green line represents the electrical noise without any light (background); the blue line represents the shot noise with only beam-1 and beam-2 (reference); the red line represents the shot noise with squeezed lights (target).

The spectrum of the raw-signal measurement is shown in {Fig.~\ref{fig:broad_spectra}(a)}. The peak at 10 MHz is the phase-insensitive heterodyne signal of two beams, while the peaks at $10\pm 3.11$ MHz are caused by the phase modulation of beam-1. The spectrum from 9 to 12 MHz is suppressed due to the 10-MHz band-stop filter. The small peak at 10 MHz in the electrical noise trace may be attributed to unintentional crosstalk from the 10-MHz electrical signals used in our control system. On the other hand, the spectrum of the demodulated-signal measurement is shown in {Fig.~\ref{fig:broad_spectra}(b)}. The peak at $3.11$ MHz is caused by the phase modulation of beam-1. The low-frequency signal below $\sim$1 MHz is suppressed by the $1.2$-MHz high-pass filter.

As can be seen from Fig.~\ref{fig:broad_spectra}, the noise reduction by squeezed light (the noise-level difference between blue and red lines) is only about 2 dB, which is below the 3-dB noise penalty. This is because the power of the pure shot noise and the electrical noise is comparable in our experimental setting. In general, we can make the influence of the shot noise sufficiently larger than that of the electrical noise by increasing the beam power as long as the detector is not saturated. However, if we increase the beam power in our setup, the laser intrinsic phase noise starts to dominate the entire signal. This is because laser intrinsic phase noise power (originates from the $E_1E_2$ term in Eq. (3) of the main text) is proportional to the square of the beam power while shot noise power (originates from the $E_2$ and $E_1$ terms) is only proportional to the beam power. Hence, the shot-noise reduction is more difficult to observe as the beam power increases; this is why the beam power is intentionally limited in our setup. Note that such laser intrinsic phase noise often survives even in the balanced detection scheme due to the independent noise source on each beam's path and the path-length difference between the two beams. To evaluate the pure noise reduction of the shot noise itself, the power of the background electrical noise is subtracted in Figs. 3 and 4 in the main text.

\bmsection{Appendix D. Demodulated spectra without cross-spectrum}

\begin{figure}[htbp]
    \centering
    \includegraphics[width=12cm]{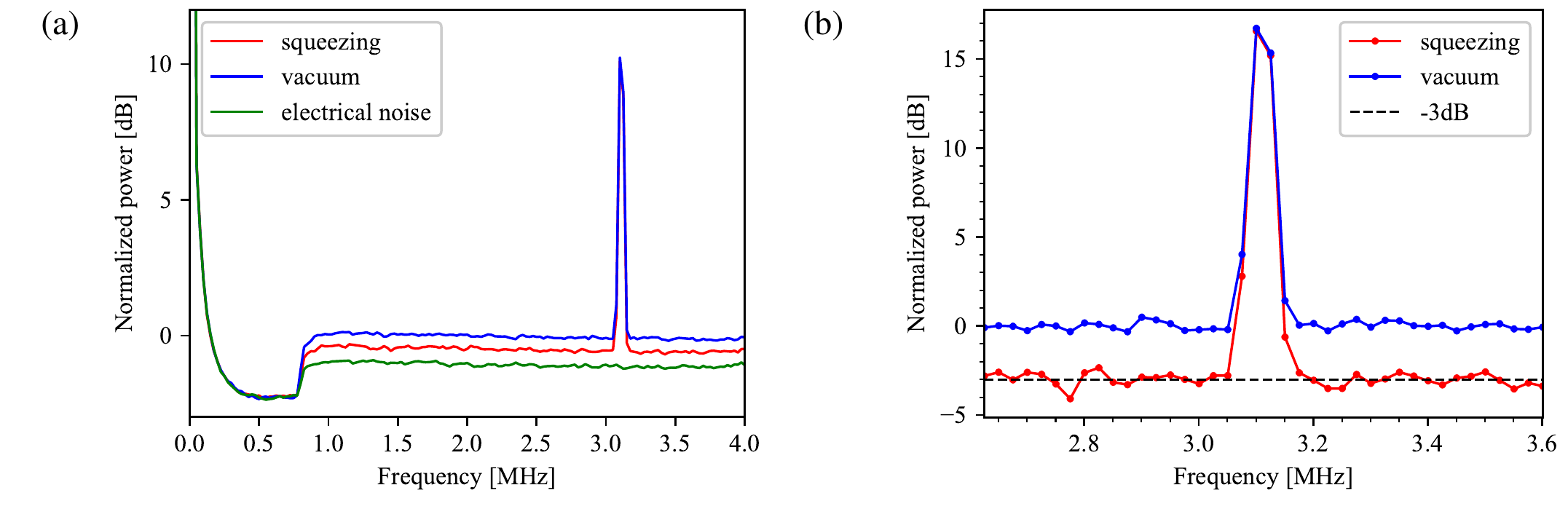}
    \caption{The demodulated spectra without cross-spectrum. (a) A broad 
    spectrum in the demodulated-signal measurement. The spectrum is normalized by the power of the vacuum averaging from 1 MHz to 3 MHz. (b) A spectrum in the demodulated-signal measurement with the power of the electrical noise subtracted and without cross-spectrum. The spectra have a peak at $3.11$ MHz caused by phase modulation of beam-1. The spectra are normalized with the averaged value of the vacuum from $3.11-0.5$ MHz to $3.11-0.05$ MHz and from $3.11+0.05$ MHz to $3.11+0.5$ MHz.}
    \label{fig:spectra_without_cross}
\end{figure}

\noindent

In the main text, we only showed the cross-spectrum for the demodulated measurement of the phase-insensitive heterodyne detection. For reference, here we plot the corresponding results without taking the cross-spectrum. Figure~\ref{fig:spectra_without_cross}(a) shows the initial wide-frequency-range spectrum without taking the cross-spectrum.
Figure~\ref{fig:spectra_without_cross}(b) then shows the narrow-frequency-range post-processed one without taking the cross-spectrum. In this case, the shot-noise level with squeezed light is 2.77(5) dB lower than that with a vacuum. Here, we average the frequency range of $\pm0.5$ MHz around $3.11$ MHz after excluding that of $\pm0.05$ MHz. This result shows that the noise reduction is smaller than that of the cross-spectrum in Fig.4 in the main text and below the 3-dB noise penalty. This difference is attributed to the electrical noise after the power splitter, which can be successfully removed by taking the cross-spectrum.

\bmsection{Appendix E. Preliminary evaluation of our OPOs}

\begin{figure}[htbp]
    \centering
    \includegraphics[width=12cm]{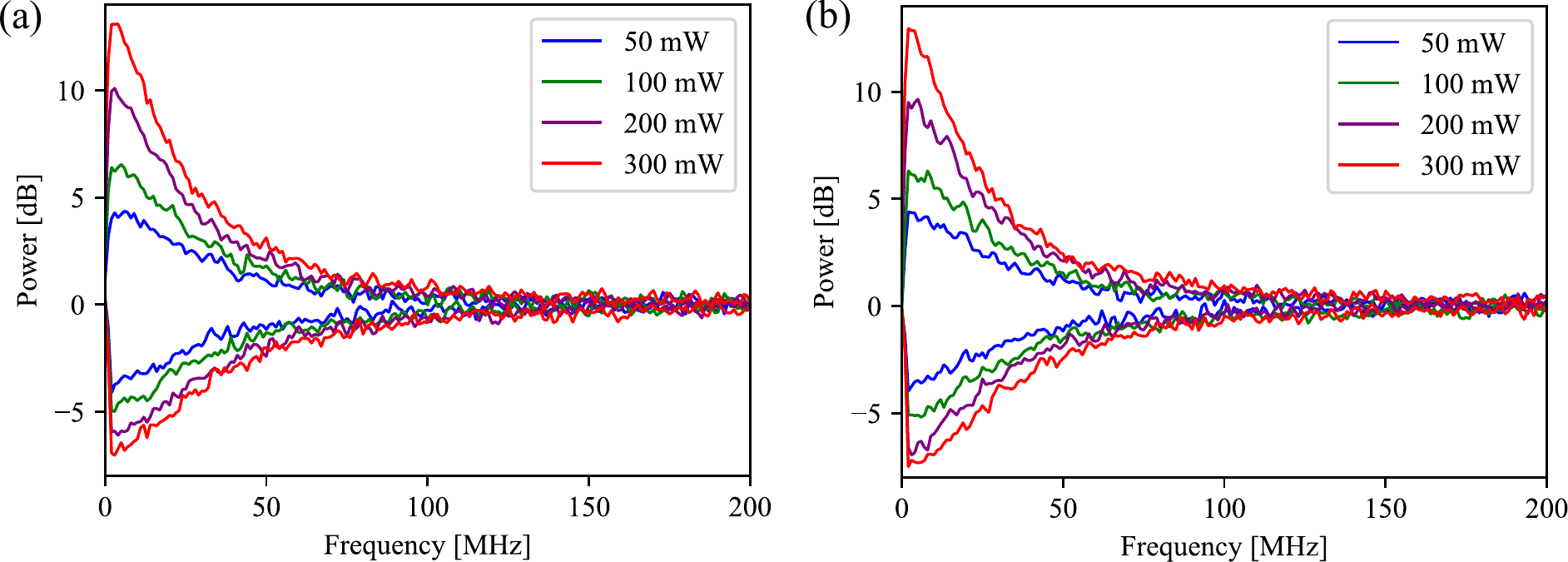}
    \caption{Squeezing spectrum of our OPOs. The horizontal axis represents frequency and the vertical axis represents the noise power, which is normalized to the shot noise level. (a) The squeezing spectrum of OPO-1. (b) The squeezing spectrum of OPO-2.}
    \label{fig:SqueezingSpectrum}
\end{figure}
Figure~\ref{fig:SqueezingSpectrum} shows the results for the preliminary evaluation of squeezed states generated from OPO-1 (Fig.~\ref{fig:SqueezingSpectrum}(a)) and OPO-2 (Fig.~\ref{fig:SqueezingSpectrum}(b)). Here, we set the power of pump beams to 50 mW (blue line), 100 mW (green line), 200 mW (purple line), and 300 mW (red line). With each power, we measure the squeezed and anti-squeezed quadratures and plot their spectrums. We can see that the squeezing level improves as the pump power increases within this power range, as expected for the OPOs operated far below the oscillator threshold ($\sim$600 mW). Note that this data was taken in a slightly different configuration from the beat-note detection, and thus we could inject higher pump power than the case in the beat-note detection.

\bmsection{Appendix F. A more general expression of shot noise in our approach}

Here, we derive Eq.~\eqref{eq:output} explicitly and without assuming that $\theta(t)$ is small. Beam-1 and beam-2 after being coupled with squeezed vacua at the high-reflectivity beam splitters can be described as
\begin{align}
    \hat{E}_1(t)&=E_1e^{i[\omega_0t+\theta_1(t)]}
    +(\hat{a}_1^{\omega_0+\Omega}(t)-i\hat{a}_2^{\omega_0+\Omega}(t))e^{i[(\omega_0+\Omega)t+\theta_\mathrm{a}]},\\
    \hat{E}_2(t)&=E_2e^{i[(\omega_0+\Omega)t+\theta_2(t)]}+(\hat{b}_1^{\omega_0}(t)-i\hat{b}_2^{\omega_0}(t))e^{i[\omega_0t+\theta_\mathrm{b}]},
\end{align}
respectively. Here, $\theta_1(t)$ and $\theta_2(t)$ are the original phases of beam-1 and beam-2 that contain the objective signal. $\theta_\mathrm{a}$ and $\theta_\mathrm{b}$ denote the squeezing angles that are controllable.
With the beam-splitter operation $\hat{E}_\pm(t)=\left[\hat{E}_1(t)\pm\hat{E}_2(t)\right]/\sqrt{2}$, the differential photocurrent signal is described as
\begin{align}
    \Delta\hat{P}(t)=&\hat{E}^\dagger_+(t)\hat{E}_+(t)-\hat{E}^\dagger_-(t)\hat{E}_-(t)\nonumber \\
    =&\hat{E}_1^\dagger(t)\hat{E}_2(t)+\hat{E}_2^\dagger(t)\hat{E}_1(t)\nonumber\\
    =&2E_1E_2\cos\left(\Omega t+\theta_2(t)-\theta_1(t)\right)\nonumber \\
    &+2E_2\left[\hat{a}_1^{\omega_0+\Omega}(t)\cos\left(\theta_2(t)-\theta_\mathrm{a}\right)-\hat{a}_2^{\omega_0+\Omega}(t)\sin\left(\theta_2(t)-\theta_\mathrm{a}\right)\right]\nonumber\\
    &+2E_1\left[\hat{b}_1^{\omega_0}(t)\cos\left(\theta_1(t)-\theta_\mathrm{b}\right)-\hat{b}_2^{\omega_0}(t)\sin\left(\theta_1(t)-\theta_\mathrm{b}\right)\right].\label{eq:output_detail}
\end{align}
By phase-locking the squeezing angle, $\theta_\mathrm{a}$ ($\theta_\mathrm{b}$) follows $\theta_\mathrm{2}(t)$ ($\theta_\mathrm{1}(t)$), and thus $|\theta_\mathrm{2}(t)-\theta_\mathrm{a}|$ ($|\theta_\mathrm{1}(t)-\theta_\mathrm{b}|$) can be made sufficiently small that the contribution from the anti-squeezed quadrature $\hat{a}_2^{\omega_0+\Omega}(t)$ ($\hat{b}_2^{\omega_0}(t)$) is negligible. Therefore, by squeezing the quadratures $\hat{a}_1^{\omega_0+\Omega}(t)$ and $\hat{b}_1^{\omega_0}(t)$, the shot noise level can be made arbitrarily low in theory.
Equation~\eqref{eq:output} can be derived from Eq.~\eqref{eq:output_detail} 
by defining $\theta(t) = \theta_2(t)-\theta_1(t)$ and assuming
$|\theta_\mathrm{2}(t)-\theta_\mathrm{a}|\approx0$ and
$|\theta_\mathrm{1}(t)-\theta_\mathrm{b}|\approx0$.

In the actual phase-locking, $|\theta_\mathrm{2}(t)-\theta_\mathrm{a}|$ and $|\theta_\mathrm{1}(t)-\theta_\mathrm{b}|$ can fluctuate and thus the anti-squeezed quadratures can contaminate the signal. However, even when the phase fluctuation is the typical value of $1.5\tcdegree$, for example, beyond 12-dB reduction of shot noise can be still achieved, as is discussed in Ref.~\cite{ths:phase_of_sq}.
\par

\bmsection{Appendix G. Straightforward approach reducing not the entire shot noise}

Here we explain that the straightforward approach to introduce phase-squeezed light cannot reduce all of shot noise in optical phase-insensitive heterodyne detection.
This approach corresponds to the case when each of beam-1 and beam-2 is coupled with a squeezed vacuum at the \textit{same} frequency as the career, written as
\begin{align}
    \hat{E}_1(t)&=E_1e^{i\omega_0 t}+\left[\hat{a}_1^{\omega_0}(t)-i\hat{a}_2^{\omega_0}(t)\right]e^{i\omega_0 t},\\
    \hat{E}_2(t)&=E_2e^{i((\omega_0+\Omega) t+\theta(t))}+\left[\hat{b}_1^{\omega_0+\Omega}(t)-i\hat{b}_2^{\omega_0+\Omega}(t)\right]e^{i((\omega_0+\Omega) t+\theta(t))}.
\end{align}
With the beam-splitter operation $\hat{E}_\pm(t)=\left[\hat{E}_1(t)\pm\hat{E}_2(t)\right]/\sqrt{2}$, the differential photocurrent signal is described as
\begin{align}
    \Delta\hat{P}(t)=&\hat{E}^\dagger_+(t)\hat{E}_+(t)-\hat{E}^\dagger_-(t)\hat{E}_-(t)\nonumber \\
    =&2E_1E_2\cos(\Omega t+\theta(t))+2E_2\left[\hat{a}_1^{\omega_0}(t)\cos(\Omega t+\theta(t))-\hat{a}_2^{\omega_0}(t)\sin(\Omega t+\theta(t))\right]\nonumber\\
    &-2E_1\left[\hat{b}_1^{\omega_0+\Omega}(t)\cos(\Omega t+\theta(t))-\hat{b}_2^{\omega_0+\Omega}(t)\sin(\Omega t+\theta(t))\right]\label{eq:typicalsig_pre}\\
    \simeq&2E_1E_2\left[\cos\Omega t-\theta(t)\sin\Omega t\right]+2E_2\left[\hat{a}_1^{\omega_0}(t)\cos\Omega t-\hat{a}_2^{\omega_0}(t)\sin\Omega t\right]\nonumber\\
    &-2E_1\left[\hat{b}_1^{\omega_0+\Omega}(t)\cos\Omega t-\hat{b}_2^{\omega_0+\Omega}(t)\sin\Omega t\right],\label{eq:typicalsig}
\end{align}
after neglecting the second-order terms for the quantum fluctuation. 
To derive Eq.~\eqref{eq:typicalsig} from Eq.~\eqref{eq:typicalsig_pre},
we also assume $\theta(t)$ to be small for simplicity.
\par

At first glance, Eq.~\eqref{eq:typicalsig} seemingly implies that this straightforward approach can remove the phase noise of the signal completely from the following reasoning.
First, in Eq.~\eqref{eq:typicalsig}, the phase signal $\theta(t)$ is proportional to $\sin\Omega t$ and thus only the terms $\hat{a}_2^{\omega_0}(t)\sin\Omega t$ and $\hat{b}_2^{\omega_0+\Omega}(t)\sin\Omega t$ seem to contribute to the phase noise. As a result, by squeezing the terms $\hat{a}_2^{\omega_0}(t)$ and $\hat{b}_2^{\omega_0+\Omega}(t)$, the phase noise can be removed completely. 
In this case, the terms $\hat{a}_1^{\omega_0}(t)$ and $\hat{b}_1^{\omega_0+\Omega}(t)$ are anti-squeezed, but they are not likely to contribute to the phase noise because these terms are proportional to $\cos\Omega t$, which corresponds to the amplitude component of the signal.
\par

However, this is not the case. Intuitively, this is because $\hat{a}_1^{\omega_0}(t)$ and $\hat{b}_1^{\omega_0+\Omega}(t)$ have not only low-frequency components but also high-frequency components, including the components around $2\Omega$. Thus, in particular, the component of $\sin 2\Omega t$ in $\hat{a}_1^{\omega_0}(t)$ and $\hat{b}_1^{\omega_0+\Omega}(t)$ turns into the phase component $\sin \Omega t$ through the term of $\sin 2\Omega t\cos \Omega t$.
As a more precise discussion of the above, we show below that the term $\hat{a}_1^{\omega_0}(t)\cos\Omega t$ actually contributes to the phase noise. In the frequency domain, the term $\hat{a}_1^{\omega_0}(t)\cos\Omega t$ can be decomposed into
\begin{align}
    \hat{a}_1^{\omega_0}(t)\cos\Omega t=&\left[\int_{-\infty}^{\infty}\frac{\mathrm{d\epsilon}}{2\pi}\hat{a}_1^{\omega_0}(\epsilon)e^{-i\epsilon t}\right]\cos\Omega t\nonumber\\
    =&\left[\int_{3\Omega}^{\infty}\frac{\mathrm{d\epsilon}}{2\pi}\hat{a}_1^{\omega_0}(\epsilon)e^{-i\epsilon t}\right]\cos\Omega t+\left[\int_{\Omega}^{3\Omega}\frac{\mathrm{d\epsilon}}{2\pi}\hat{a}_1^{\omega_0}(\epsilon)e^{-i\epsilon t}\right]\cos\Omega t\nonumber\\
    &+\left[\int_{-\Omega}^{\Omega}\frac{\mathrm{d\epsilon}}{2\pi}\hat{a}_1^{\omega_0}(\epsilon)e^{-i\epsilon t}\right]\cos\Omega t+\left[\int_{-3\Omega}^{\Omega}\frac{\mathrm{d\epsilon}}{2\pi}\hat{a}_1^{\omega_0}(\epsilon)e^{-i\epsilon t}\right]\cos\Omega t\nonumber\\
    &+\left[\int_{-\infty}^{-3\Omega}\frac{\mathrm{d\epsilon}}{2\pi}\hat{a}_1^{\omega_0}(\epsilon)e^{-i\epsilon t}\right]\cos\Omega t.\label{eq:ft}
\end{align}
Here, the first, third, and last terms in the right-hand side of Eq.~\eqref{eq:ft} do not contribute to the phase noise which is proportional to $\sin\Omega t$. Hence, we consider only the second and fourth terms. The second term can be calculated as
\begin{align}
    \left[\int_{\Omega}^{3\Omega}\frac{\mathrm{d\epsilon}}{2\pi}\hat{a}_1^{\omega_0}(\epsilon)e^{-i\epsilon t}\right]\cos\Omega t&=\left[\int_{-\Omega}^{\Omega}\frac{\mathrm{d\epsilon}}{2\pi}\hat{a}_1^{\omega_0}(\epsilon-2\Omega)e^{-i(\epsilon-2\Omega) t}\right]\cos\Omega t\nonumber\\
    &=\int_{-\Omega}^{\Omega}\frac{\mathrm{d\epsilon}}{2\pi}\hat{a}_1^{\omega_0}(\epsilon-2\Omega)e^{-i\epsilon t}\left(\cos2\Omega t+i\sin2\Omega t\right)\cos\Omega t\nonumber\\
    &=\int_{-\Omega}^{\Omega}\frac{\mathrm{d\epsilon}}{2\pi}\hat{a}_1^{\omega_0}(\epsilon-2\Omega)e^{-i\epsilon t}\left[\frac{1}{2}\left(\cos\Omega t+\cos3\Omega t\right)+\frac{i}{2}\left(\sin\Omega t+\sin3\Omega t\right)\right]. \label{eq:FTpartial}
\end{align}
In the same way, the fourth term can be calculated as
\begin{equation}
    \left[\int_{-3\Omega}^{\Omega}\frac{\mathrm{d\epsilon}}{2\pi}\hat{a}_1^{\omega_0}(\epsilon)e^{-i\epsilon t}\right]\cos\Omega t=\int_{-\Omega}^{\Omega}\frac{\mathrm{d\epsilon}}{2\pi}\hat{a}_1^{\omega_0}(\epsilon+2\Omega)e^{-i\epsilon t}\left[\frac{1}{2}\left(\cos\Omega t+\cos3\Omega t\right)-\frac{i}{2}\left(\sin\Omega t+\sin3\Omega t\right)\right]. \label{eq:FTpartial2}
\end{equation}
From Eqs.~\eqref{eq:FTpartial}~and~\eqref{eq:FTpartial2}, the $\sin\Omega t$ component in Eq.~\eqref{eq:ft} can be summarized as
\begin{equation}
    \frac{i}{2}\int_{-\Omega}^{\Omega}\frac{\mathrm{d\epsilon}}{2\pi}\left[\hat{a}_1^{\omega_0}(\epsilon-2\Omega)-\hat{a}_1^{\omega_0}(\epsilon+2\Omega)\right]e^{-i\epsilon t}\sin\Omega t. \label{eq:PhaseNoise}
\end{equation}
This means that the term $\hat{a}_1^{\omega_0}(t)\cos\Omega t$ contains the $\sin\Omega t$ component and thus contributes to the phase noise.
Notably, if we try to reduce the phase noise in Eq.~\eqref{eq:typicalsig} by squeezing $\hat{a}_2^{\omega_0}(t)$, $\hat{a}_1^{\omega_0}(t)$ is anti-squeezed,
and thus $\hat{a}_1^{\omega_0}(\epsilon-2\Omega)$ and $\hat{a}_1^{\omega_0}(\epsilon+2\Omega)$ in Eq.~\eqref{eq:PhaseNoise} are anti-squeezed independently and contaminate the phase noise instead.
Hence, we cannot reduce shot noise from all relevant bands in this straightforward approach.

\bmsection{Funding}
Japan Science and Technology Agency (JPMJPF2221, JPMJFR223R, JPMJPR2254); Japan Society for the Promotion of Science (20H01833, 21K13933, 21K18593).

\bmsection{Acknowledgments}
S. T. acknowledges supports from MEXT Leading Initiative for Excellent Young Researchers, the Canon Foundation, UTokyo Foundation, and donations from Nichia Corporation.
K. A. acknowledges a financial support from Forefront Physics and Mathematics Program to Drive Transformation (FoPM). M. E. was supported by the Research Foundation for Opto-Science and Technology. The authors thank Takahiro Mitani for the careful proofreading of the manuscript.

\bmsection{Disclosures}
The authors declare no conflicts of interest.

\bmsection{Data Availability}
Data underlying the results presented in this paper are not publicly available at this time but may be obtained from the authors upon reasonable request.


\end{backmatter}








\end{document}